# On the Dynamics of Generalized Coherent States.
## II. Classical Equations of Motions

B.A. Nikolov, D.A. Trifonov

**Abstract.** Using the Klauder approach the stable evolution of generalized coherent states (GCS) for some groups ($SU(2)$, $SU(1,1)$ and $SU(N)$) is considered and it is shown that one and the same classical solution $z(t)$ can correctly characterize the quantum evolution of many different (in general nonequivalent) systems. As examples some concrete systems are treated in greater detail: it is obtained that the nonstationary systems of the singular oscillator, of the particle motion in a magnetic field, and of the oscillator with friction all have stable $SU(1,1)$ GCS whose quantum evolution is determined by one and the same classical function $z(t)$. The physical properties of the constructed $SU(1,1)$ GCS are discussed and it is shown particularly that in the case of discrete series $D_k^{(+)}$ they are those states for which the quantum mean values coincide with the statistical ones for an oscillator in a thermostat.

## 1 Introduction

In the previous paper [1] (to which we shall refer as Ref. I) the exact and stable time evolution of generalized coherent states (GCS) was discussed and a method for constructing GCS related to any Lie group and any quantum system was described. The aim of the present paper is to study the dynamics of GCS on the example of some concrete groups. The exact and stable evolution of $SU(1,1)$ GCS is obtained for nonstationary oscillator, for a motion of a particle in time-dependent magnetic field, and for a singular oscillator with time-dependent friction. The physical properties of the constructed GCS are discussed and in particular it is shown that the GCS related to discrete series of unitary irreducible representations (UIR) of $SU(1,1)$ are exactly those quantum states for which the quantum mean values of observables are equal to statistical averages. Using the Klauder approach we obtain the classical Euler equations ($|\Phi\rangle$ being the GCS I.(25)[1], $z_j$ - local coordinates for $X = G/K$)

$$\frac{d}{dt}\frac{\partial \mathcal{L}}{\partial \dot{z}_j} - \frac{\partial \mathcal{L}}{\partial z_j} = 0, \tag{1}$$

$$\mathcal{L} = i\langle \Phi_z | d/dt | \Phi_z\rangle - \langle \Phi_z | H | \Phi_z |\rangle$$

---
[1] I.(25) = Eq. (25) of Ref. I. (Note added in the e-print quant-ph/0407261).

as independent of the type of representation. This means that one and the same classical solution of Euler equations correctly determines the quantum evolution of many different initial states (distinguished by different values of some conserved quantities) and different Hamiltonians.

## 2 Glauber Coherent States

These states are related to the projective unitary representation of the phase space translation group. The representation operators form the so-called Heisenberg-Weyl group, which is generated by lowering and raising operators $a$, $a^\dagger$ and identity operator $\hat{I}$, $[a, a^\dagger] = \hat{I}$. The overcomplete family of states (OFS) is given by (see I.(3), I.(4)):

$$|(\beta, z, z^*)\rangle = \exp(i\beta I + za^\dagger - z^*a)|\Phi_0\rangle, \quad z \in C. \tag{2}$$

The stationary group $K$ (I, Sec. 2) is trivially generated by $\hat{I}$. The quotient space $G_W/K$ ($G_W$ - Heisenberg-Weyl group) is isomorphic to the complex plane $C$. Choosing the cross-section $s : (z, z^*) \to (0, z, z^*)$ and $|\Phi_0\rangle = |0\rangle$ ($a|0\rangle = 0$) we obtain the system of Glauber CS (I, Refs. 1,2). Another choice, say $s' : (z, z^*) \to (\beta(z, z^*), z, z^*)$ permits to make manifest the stability of the system of CS (I, Ref. 1). The symplectic 2-form, determined by the Kähler potential $f = z^*z$ is $\omega = idz \wedge dz^*$ and therefore the equation of motion is ($\partial_z^* = \partial/\partial z_i^*$)

$$i\dot{z} = \partial_z^* \mathcal{H}. \tag{3}$$

(the complex conjugate equations will not be written down). The same equation can be derived making use of the Lagrangian $\mathcal{L} = (i/2)(\dot{z}z^* - \dot{z}^*z) - \mathcal{H}$, according to Klauder method (I, Sec. 3).

For the most general Hamiltonian I.(19) which preserves the OFS (2) stable, Eq. (3) assumes the form

$$i\dot{z} = \omega(t)z + F(t). \tag{4}$$

The solution of this equation can be explicitly written:

$$z(t) = \left(z - i \int F(t)dt\right) \exp\left(-i \int \omega(t)dt\right), \quad z = z(0).$$

We see that the phase space trajectory of the considered system is a superposition of a translation and a rotation. This result simply reflects the fact that the Hamiltonian I.(19) belongs to the (projective) representation of Lie algebra $e(2)$ of Euclidean group $E(2)$.

## 3 SU(2) Coherent States

The generators of the group $SU(2)$ are $J_i$, $[J_i, J_j] = i\epsilon_{ijk}J_k$, and the Casimir operator is equal to $J^2 = J_1^2 + J_2^2 + J_3^2 = j(j+1)$, $j = 1/2, 1, 3/2, \ldots$. The three parameters usually used are the Euler angles $(\phi, \theta, \psi)$. As in Ref. 12 of [1] we choose $|\Phi_0\rangle = |j, -j\rangle$, where $|j, m\rangle$ are the eigenvectors of $J_3$: $J_3|j, m\rangle = m|j, m\rangle$, $m = -j, -j+1, \ldots, j$, and such a cross-section in the fibre bundle $(G, G/K, \pi)$ ($K$ being the stationary subgroup of $|j, -j\rangle$) that

$$|\Phi(\phi, \theta)\rangle = |j; z\rangle = (1 + z^*z)^{-j} \exp(zJ_+)|j, -j\rangle, \tag{5}$$

where $z = -\tan\theta/2\, e^{-i\phi}$, $J_+ = J_1 + iJ_2$.

The Hamiltonian

$$H = h_i J_i = hJ_+ + h^* J_- + h_0 J_0,$$
$$h = h_1 + ih_2, \quad h_0 = h_3, \quad J_- = (J_+)^\dagger, \tag{6}$$

according to Malkin theorem (I, Sec. 3) should preserves all $SU(2)$ GCS (5) stable. The classical solution $z(t)$ obeys the Euler eqs. for the functional I.(25), which in this case have the form

$$i\dot{z} = (2j)^{-1}(1 + z^*z)^2 \partial\mathcal{H}/\partial z^*. \tag{7}$$

The same equation is obtained in [2], where it arises as equation of the path with the main contribution in the path integral, which expresses the transition amplitude from a GCS (5) to another one.

By means of the easily verified formulae ($\langle . \rangle_{jz} \equiv \langle z;j|(.)|j;z\rangle$)

$$\langle J_+ \rangle_{jz} = \frac{2jz^*}{1+z^*z}, \quad \langle J_- \rangle_{jz} = \frac{2jz}{1+z^*z}, \quad \langle J_0 \rangle_{jz} = -\frac{1-z^*z}{1+z^*z}$$

we can get the classical Hamiltonian

$$\mathcal{H} = j(1+z^*z)^{-1}\left(2(hz + h^*z^*) - h_0(1 - z^*z)\right). \tag{8}$$

From (7) and (8) we obtain the Euler equation

$$i\dot{z} = h^* + h_0 z - hz^2. \tag{9}$$

Now it is worth noticing the important property of Eq. (9), namely it does not depends on the representation $(j)$ of $SU(2)$. One and the same classical function $z(t)$ entirely determines the quantum evolution of all systems of GCS $((j), |j;-j\rangle)$ governed by the Hamiltonian (8). Taking different representations for the angular momentum $J_i$ we get different (and non-equivalent) quantum systems whose dynamics is exactly determined by the same classical trajectory in phase space $X = G/K$.

As we have stated in the previous paper I, any OFS, in particular the system of GCS (5), can be realized in the Hilbert space $\boldsymbol{H}$ of solutions by means of Eq. I.(15) provided the generators $L_a$ are expressed in terms of $a$, $a^\dagger$. The most natural representation in $\boldsymbol{H}$ for the angular momentum $J_i$, $J^2 = j(j+1)$, is that of $2j+1$ numbers of operators $a_i$, $a_i^\dagger$ (I, Ref. 22), $i = 1, 2, \ldots, 2j+1$,

$$J_k = a_i^\dagger (\hat{J}_k)_{il} a_l, \tag{10}$$

where $\hat{J}_k$ are $(2j+1)$-dimensional matrices. Formula (10) generalizes the Schwinger representation $(j = 1/2)$ [3] to the case of arbitrary $j$. In [4] another angular momentum boson representation is constructed for arbitrary $j$ in terms of $2s+1$ pairs $a_i, a_i^\dagger$, $i = -s, -s+1, \ldots, s$, $s = 1/2, 3/2, \ldots$. The natural representation (10) is irreducible in Fock space, spanned by the vectors

$$|n\rangle = \sum_{i=1}^{2j+1} (n_i!)^{-1/2} (a_i^\dagger)^{n_i} |0\rangle \tag{11}$$

with a fixed number $n = \sum_i n_i = 2j$.

The one-dimensional systems are in some sense exceptional but it is still possible to express any UIR of $SU(2)$ in terms of one pair $a, a^\dagger$ (I, Ref. 24) in the Fock space of vectors $|m\rangle = (m!)^{-1/2}(a^\dagger)^m|0\rangle$, $m = 0, 1, \ldots, 2j$ ($|0\rangle = |j; -j\rangle$, $|m\rangle = |j, -j + m\rangle$) and [2]

$$V^{(j)}|m\rangle = (m!)^{-1/2}(1 + z^*z)^{-j}(1 + za^\dagger)^{2j-m}(a^\dagger - z)^m|0\rangle.$$

Putting $J_k$ from Eq. (10) into (6) we obtain Hamiltonians for different quadratic quantum systems, all having the property their dynamics determined by one classical function $z(t) = -\tan\theta(t)/2 \exp(-i\phi(t))$, $\theta$ and $\phi$ being the Euler angles for the $SU(2)$ rotations. The $N$-dimensional oscillator is one of the simplest such systems. In this case $z(t) = z(0)\exp(i\omega t)$.

Let us note that the $SU(2)$ CS constructed from Schwinger operators $J_- = a^\dagger b$, $J_+ = b^\dagger a$, $J_0 = (1/2)(a^\dagger a - b^\dagger b$, namely

$$|N; z\rangle = (1 + z^*z)^{-N/2} \sum_{n=0}^{N} z^n |N; n\rangle, \tag{12}$$

$$|N; n\rangle = (n!(N-n)!)^{-1/2}(a^\dagger)^n(b^\dagger)^{N-n}|0; 0\rangle,$$

$$a|0;0\rangle = b|0;0\rangle = 0, \quad N = 0, 1, 2, \ldots$$

represent an OFS only in the $(N+1)$-dimensional subspace of Hilbert space $\boldsymbol{H} \times \boldsymbol{H}$.

It is possible however, using GCS (12), to construct wave packets, which form OFS in the whole space $\boldsymbol{H} \times \boldsymbol{H}$. Indeed, the simple calculation provides the relation ($\alpha \in C$)

$$\sum_{N=1}^{\infty}(N!)^{-1/2}\alpha^N|N; z\rangle = \exp\left(\alpha(za^\dagger + b^\dagger)(1 + z^*z)^{-1/2}\right)|0;0\rangle. \tag{13}$$

Introducing the new lowering and raising operators

$$\begin{aligned} A_z &= (1 + z^*z)^{-1/2}(z^*a + b), \\ A_z^\dagger &= (1 + z^*z)^{-1/2}(za^\dagger + b^\dagger) \end{aligned} \tag{14}$$

we see that the wave packets (13) are (apart from a normalization constant) just the Glauber CS for $A_z$, $A_z^\dagger$.

From the other hand the states (13) obviously are tensor products of the type $|\lambda\rangle|\mu\rangle$, where $\lambda = \alpha z(1 + z^*z)^{-1/2}$, $\mu = \alpha(1 + z^*z)^{-1/2}$ ($|\lambda\rangle$, $|\mu\rangle$ being one mode Glauber CS), and consequently they form an OFS in $\boldsymbol{H} \times \boldsymbol{H}$. These wave packets have been called "oscillator like CS for rotation group"[5].

From the known result for the Heisenberg-Weyl group $G_W$ we obtain that the most general Hamiltonian which preserves the OFS (13) stable has the form $H = \omega A_z^\dagger A_z + f^*A_z + fA_z^\dagger + \beta$, i.e. the stable OFS (13) admits wider class of Hamiltonians than the set (12). The Euler equations for such systems in terms of parameters $\lambda$, $\mu$ have the form (4).

The physical properties of $SU(2)$ CS will not be discussed here since they have been thoroughly examined (see I, Refs. 4,10,12).

---

[2] The states $(m!)^{-1/2}(1+z^*z)^{-j}(1+za^\dagger)^{2j-m}(a^\dagger-z)^m|0\rangle \equiv |z; 2j, m\rangle$ are not normalized. (Note added).

# 4 SU(1,1) Coherent States

The UIR of $SU*1,1)$ are generated by the operators $K_i$ ($i = 1, 2, 3$) with the commutation relations

$$[K_1, K_2] = -iK_3, \quad [K_2, K_3] = iK_1. \quad [K_3, K_1] = iK_2, \tag{15}$$

and the Casimir operator is $K^2 = K_3^2 - K_1^2 - K_2^2 = k(k-1)$, $k > 0$. (We restrict ourselves with the discrete series $V_k^{(+)}$.) All UIR have been described by Bargmann [6].

As in the previous section let us study first the stable evolution of $SU(1,1)$ CS for the Hamiltonian

$$H = h_0 K_3 + h_1 K_1 + h_2 K_2 = h_0 K_0 + h^* K_- + h K_+ \tag{16}$$

by making use of Klauder approach. The canonical basis in Hilbert space $\boldsymbol{H}$ is $|k; m\rangle$, $m = 0, 1, 2, \ldots$, $K_0 |k; m\rangle = (k + m)|k; m\rangle$. As in I, Ref. 12, we choose the lowest weight vector $|k; 0\rangle$ as a fiducial one. Its stationary group is the subgroup of rotations around the third axis. The system of GCS is written in the form

$$|k; z\rangle = (1 - z^* z)^k \exp(z K_+) |k; 0\rangle, \tag{17}$$

where $z = -\tanh \tau/2 \exp(-i\phi)$, $\tau$ and $\phi$ being the Euler angles for $SU(1,1)$. Using the formulae

$$\langle K_+ \rangle_{kz} = \frac{2kz^*}{1 - z^* z}, \quad \langle K_- \rangle_{kz} = \frac{2kz}{1 - z^* z}, \quad \langle K_0 \rangle_{kz} = k \frac{1 + z^* z}{1 - z^* z}$$

we obtain the Euler equations for the functional I.(25)

$$i\dot{z} = (2k)^{-1}(1 - z^* z)^2 \, \partial \mathcal{H}/\partial z^*, \tag{18}$$

where for the case of linear Hamiltonian (16)

$$\mathcal{H} = k \left(h_0(1 + z^* z) + 2hz^* + 2h^* z\right)(1 - z^* z)^{-1}, \tag{19}$$

and therefore Eq. (18) assumes the form

$$i\dot{z} = h^* z^2 + h_0 z + h, \tag{20}$$

quite similar to the analogical one (9) for the $SU(2)$ group, and again independent of the representation $V_k^{(+)}$.

The phase space corresponding to the UIR $V_k^{(k)}$ is the Lobachevsky plane represented by disc $D : |z| < 1$ with the Kähler potential (see I.(28)) $f = \ln(1 - z^* z)^{-2k}$. Hence the symplectic form is

$$\omega = 2ik(1 - z^* z)^{-2} dz \wedge dz^*,$$

and $g = (1 - z^* z)^2 / 2ik$ (see I.(28)). Thus we again arrive to Eq. (18) by means of the Lie bracket I.(29).

Next we shall construct the $SU(1,1)$ CS for some concrete quantum systems and investigate some of their properties.

## 4.1 Nonstationary Quantum Oscillator with Friction

The Hamiltonian of this system is

$$H = \frac{1}{2}\left(p^2 + \omega(t)^2 q^2\right) + \frac{1}{2}b(t)(qp + pq), \quad [q,p] = i. \tag{21}$$

Using the following representation of the Lie algebra $su(1,1)$ [7]

$$K_1 = \frac{1}{4}(p^2 - q^2), \quad K_2 = \frac{1}{4}(pq + qp), \quad K_3 = \frac{1}{4}(p^2 + q^2) \tag{22}$$

we can write down (21) as a linear combination of generators (22),

$$H = (1 - \omega^2)K_1 + 2bK_2 + (1 + \omega^2)K_3 = h_0 K_0 + h K_+ + h^* K_-, \tag{23}$$

where $h = (1 - \omega^2)/2 - ib$, $h_0 = 1 + \omega^2$, $K_\pm = K_1 \pm iK_2$, $K_0 = K_3$.

Let us find the $SU(1,1)$ CS for UIR, generated by the operators (22). The Casimir operator is $K^2 = -3/16$, hence there are two UIR belonging to discrete series $V_k^{(+)}$: $k = 1/4$ and $k = 3/4$. The basis for $k = 1/4$ ($k = 3/4$) is formed by the even (odd) states of Fock space, corresponding to the eigenvalue $+1$ ($-1$) of parity operator (that is why we shall denote $k = +$ ($k = -$)). The Fock space is generated by the creation operator $a^\dagger = (p + iq)/\sqrt{2}$, then $|+; m\rangle = |2m\rangle$, $|-; m\rangle = |2m + 1\rangle$, where $|n\rangle = a^n|0\rangle/\sqrt{n!}$. The corresponding $SU(1,1)$ CS are

$$\begin{aligned}|+; z\rangle &= (1 - z^* z)^{1/4} \exp\left(z(a^\dagger)^2/2\right) |0\rangle, \\ |-; z\rangle &= (1 - z^* z)^{3/4} \exp\left(z(a^\dagger)^2/2\right) |1\rangle\end{aligned} \tag{24}$$

or, in coordinate representation ($N_\pm$ - normalization constants, depending on $z$),

$$\langle x|\pm; z\rangle = N_\pm\, x^{(1\mp 1)/2} e^{-ax^2}, \quad a = \frac{1}{2}\frac{1 + z}{1 - z}. \tag{25}$$

The related probability densities

$$\begin{aligned}w_\pm(x) &= \pi^{-1/2}(2x^2)^{(1\mp 1)/2}\lambda^{1\mp 1/2} e^{-\lambda x^2}, \\ \lambda &= (1 - z^* z)|1 - z|^{-2}\end{aligned} \tag{26}$$

describe distributions of Gaussian type having maximums at the points $x_\pm$, $x_\pm^2 = (1\mp 1)/2\lambda$ and width (distance between the extreme inflex points) also proportional to $\lambda^{-1/2}$. Note to the point that the width of the Gaussian distributions, corresponding to Glauber CS $|z\rangle$, does not depend on the label $z$.

The computation of the mean values of operators $q^2$ and $p^2$ yields

$$\begin{aligned}\langle q^2\rangle_{kz} &= 2k(1 - z^* z)^{-1}|1 - z|^2 = 2k\lambda, \\ \langle p^2\rangle_{kz} &= 2k(1 - z^* z)^{-1}|1 + z|^2.\end{aligned} \tag{27}$$

Multiplying these quantities one obtains the Heisenberg uncertainty product (since $\langle q \rangle_{kz} = 0$, $\langle p \rangle_{kz} = 0$):

$$(\Delta q)_{kz}^2 (\Delta p)_{kz}^2 = \langle q^2 \rangle_{kz} \langle p^2 \rangle_{kz} = 4k^2 \frac{1 + r^4 - 2r^2 \cos(2\theta)}{1 - r^2} \geq 4k^2, \qquad (28)$$

where $r = |z|$, $\theta = \arg z$. The equality holds iff $z$ is real nonnegative number: $z = r \geq 0$. Hence the only minimum uncertainty states (MUS) from the considered GCS system are $|+; r\rangle = (1 - r^2)^{1/4} \exp\left(r a^{\dagger 2}/2\right) |0\rangle$, $r > 0$. They are unitarily equivalent to the Glauber CS in accordance with the result obtained by Stoler [8].

Turning to the time evolution and observing that the Hamiltonian (21) obeys the Malkin theorem conditions (I, Sec. 3) we conclude that the OFS (24) are stable and in every moment $t$ the $SU(1,1)$ CS are determined by the same formulae (24), where $z = z(t)$ is a solution of the Euler equations (20).

On the other hand one can obtain the exact solution making use of the integrals of motion method (I, Ref. 11). Comparing the two solutions it is easy to express $z(t)$ as a fractional linear transformation:

$$z(t) = \frac{a(t) z + c(t)}{c^*(t) z + a^*(t)}, \quad a = (\rho^2 + 1) e^{i\gamma}, \; c = (\rho^2 - 1) e^{-i\gamma}, \qquad (29)$$

where $\epsilon = \rho e^{i\gamma}$ is a solution of the following classical equation:

$$\ddot{\epsilon} + \Omega^2 \epsilon, \quad \Omega^2 = \omega^2 - b^2 - \dot{b}, \; \rho^2 \dot{\gamma} = 1. \qquad (30)$$

Eq. (29) explicitly shows that the time evolution is represented as a $SU(1,1)$ transformation in the phase space $D$: $|z| \leq 1$.

## 4.2 Generalized Singular Oscillator

The Hamiltonian of this system contains a singular term which may serve as a model potential describing interaction between particles (see I, Refs. 9, 11):

$$H = \frac{1}{2} \left( p^2 + \omega(t)^2 q^2 \right) + \frac{1}{2} b(t)(qp + pq) + g/q^2. \qquad (31)$$

We suppose that $0 < q < \infty$ since the singular potential prohibits transition from $(\infty, 0)$ to $(-\infty, 0)$. Further we restrict ourselves with sufficiently large values of the interaction constant $g$ ($g > -1/8$) because otherwise a collapse would be possible (I, Ref. 11).

We shall use the method of integrals of motion. The latter are determined in the form [9]:

$$M_- = \frac{1}{2} \left( a^2 + g \epsilon^2(t) q^{-2} \right), \quad M_+ = (M_-)^\dagger, \quad M_0 = \frac{1}{2}[M_-, M_+],,$$
$$a = \frac{i}{\sqrt{2}}(\epsilon p + (\epsilon b - \dot{\epsilon}) q), \quad [q, p] = i, \qquad (32)$$

where $\epsilon$ is a solution of Eq. (30). The operators (32) form a representation of the Lie algebra $su(1,1)$, $[M_0, M_\pm] = \pm M_\pm$, and commute with the Schrödinger operator $D_S = i\partial/\partial t - H$. Thus we have a realization of the dynamical Lie algebra [7] of the system under consideration. The UIR of the group $SU(1,1)$, constructed by means of (32) are labeled by the number $k = (d+1)/2$ where $d = (1/2)(1+8g)^{1/2} > 0$, i.e. the operators (32) generate UIR from the discrete series $V_k^{(+)}$. Making use of the eigenvectors of $M_0$ [9]

$$\langle x|k;n\rangle = \left(2\epsilon^{-2d-2}\Gamma(n+1)/G(n+d+1)\right)^{1/2} x^{d+1/2} \times$$
$$\exp\left(-2in\gamma + (i/2)(\dot{\epsilon}/\epsilon - b)x^2\right) L_n^d(\dot{\gamma}x^2), \quad \gamma = \arg\epsilon \tag{33}$$

we construct the $SU(1,1)$ CS for our system [3]:

$$\langle x|k;z,t\rangle = (2/\Gamma(d+1))^{1/2} \epsilon^{-2d-2} (1-z^*z)^{(d+1)/2} x^{d+1/2} \times$$
$$(1-s)^{-d-1} \exp\left(-\frac{1}{2\rho^2}\frac{1+s}{1-s}x^2\right), \tag{34}$$

where $s = z\exp(-2i\gamma)$, $|z| < 1$, $k = (d+1)/2$ (a phase factor was omitted [4]).

In the initial moment $t = 0$ Eq. (34) converse to the form (up to normalization constant)

$$\langle x|k;z,0\rangle = (1-z^*z)^{(d+1)/2} x^{d+1/2}(1-z)^{-d-1} \exp\left(-\frac{1}{2}\frac{1+z}{1-z}x^2\right), \tag{35}$$

where we have put $\dot{\gamma}(0) = 1$, $\gamma(0) = 0$. The similarity with Eq. (25) is obvious. In fact formula (35) includes (25) as a special case if we put there $d = 2k-1$, $k = 1/4, 3/4$. Moreover the wave function (34) can be written in the form (35) with $z$ replaced by

$$z(t) = \frac{az+c}{c^*z+a^*}, \quad a = (1+\rho^2)e^{-i\gamma}, \quad ca = (1-\rho^2)e^{i\gamma}. \tag{36}$$

The latter expression essentially coincides with (29), i.e., the addition of the singular term $g/q^2$ in the Hamiltonian does not exert influence upon the time evolution in phase space $D$.

The dynamical symmetry group of the quantum oscillator (21) is obviously more large than $SU(1,1)$ (it must include transformations mixing states with different parity). The addition of the singular term to (21) reduces the symmetry to $SU(1,1)$, i.e., this group is a dynamical symmetry group of the singular oscillator. The trajectory in phase space however remains unchanged. Thus the motion in phase space $D$ is not obliged to reflect the dynamics of corresponding classical system. Indeed, the quantum mean value of coordinate operator $q^2 = 2M_0\epsilon^*\epsilon - \epsilon^2 M_+ - \epsilon^{*2} M_-$,

$$\langle q^2 \rangle_{kz} = \frac{2k}{1-z^*z}\left(\epsilon^*\epsilon(1+z^*z) - z^*\epsilon^2 - z\epsilon^{*2}\right),$$

varies in stationary case ($\epsilon = \Omega^{-1/2}\exp(i\Omega t)$, $\Omega = $ const) like the elongation of the usual harmonic oscillator. One can conclude also that the $SU(1,1)$ CS (34), (35) are not close to the classical states.

---

[3] $L_n^d(x)$ are generalized Laguerre polynomials, and $\Gamma(x)$ is Gamma-function. For a generalization and further properties of these states see e.g. quant-ph/9811081. (Note added).

[4] The omitted $z$-independent phase factor is $\exp(-ix^2(b - \dot{\rho}/2\rho))$. (Note added).

## 4.3 Motion of a Particle in a Time-Dependent Magnetic Field

Let us consider a particle with unit mass and unit charge ($m = 1 = e$) moving in a time dependent magnetic field $\mathcal{H}(t) = 2\omega(t)$. The Hamiltonian of this system is

$$H = \frac{1}{2}\left(p_x^2 + p_y^2\right) + \frac{1}{2}\omega^2(x^2 + y^2) + \omega(yp_x - xp_y). \tag{37}$$

Exact solution solutions of this problem were obtained in I, Ref. 5. Two independent integrals of motion can be constructed:

$$\begin{aligned} A &= \frac{1}{2}\left(\epsilon\left(p_x + ip_y\right) - i\dot{\epsilon}\left(y - ix\right)\right)e^{i\phi}, \\ B &= \frac{1}{2}\left(\epsilon\left(p_y + ip_x\right) - i\dot{\epsilon}\left(x - iy\right)\right)e^{-i\phi}, \end{aligned} \tag{38}$$

where $\phi = (1/2)\int dt\omega(t)$ and $\epsilon$ is the solution of Eq. (30) with $\Omega = \omega/2$. The operators $A$, $B$ obey the relations $[A, A^\dagger] = [B, B^\dagger] = 1$, $[A, B] = [A, B^\dagger] = 0$, and their eigenvalues determine the running coordinates of the wave-packet center and the coordinates of the orbital center in $(x, y)$-plane respectively. The corresponding eigenvectors are the Glauber CS for this system (I, Ref. 11).

From the operators $A$, $B$ we build the following representation of Lie algebra $su(1,1)$ (I, Refs. 7, 29):

$$K_+ = A^\dagger B^\dagger, \quad K_- = AB, \quad K_0 = (1/2)(A^\dagger A + B^\dagger B + 1). \tag{39}$$

The Casimir operator $K^2$ can be expressed by the third projection of the angular momentum $L_3 = B^\dagger B - A^\dagger A$: $K^2 = (L_3^2 - 1)/4$. Denoting the eigenvectors of Hermitian[5] operators $A^\dagger A$, $B^\dagger B$ by $|n, m; t\rangle$ we obtain that the UIR of the group $SU(1,1)$, generated by operators $A$, $B$ is $V_k^{(+)}$, $k = (N+1)/2$, $N = m - n$, and it is spanned by the vectors $|k; n\rangle = |n, N+n; t\rangle$, $n = 0, 1, 2, \ldots$.

Now it is not difficult to write down the $SU(1,1)$ CS using the explicit form of vectors $|n, m; t\rangle$, given in Ref. 11 of I (up to a phase factor):

$$\begin{aligned} \langle w|n, N+n; t\rangle &= \left(\frac{n!}{\pi(n+N)!}\right)^{1/2}\left(-ie^{-2i\gamma}\right)^n \times \\ &\quad \rho^{-N-1}r^N \exp\left(-r^2/2\rho^2\right) L_n^N\left(r^2/\rho^2\right), \end{aligned} \tag{40}$$

where

$$w = -2^{-1/2}(x + iy)e^{i\phi} = 2^{-1/2}re^{i\theta}.$$

Then by means of Eq. (17) we obtain

$$\begin{aligned} \langle w|n, N+n; t\rangle &= (\pi N!)^{-1/2}\left(2\mathrm{Re}\, a\right)^{(N+1)/2} r^N \exp(-ar^2), \\ a &= (1+s)/(2\rho(1-s)), \quad s = -ize^{-2i\gamma}, \quad k = (N+1)/2. \end{aligned} \tag{41}$$

---
[5] "Hermitean" and "Weil" changed to "Hermitian" and "Weyl". (Note added).

It is clear from the construction that these systems of GCS have an important property: all wave functions (41) with fixed $N$ belong to the eigenspace of the third projection of angular momentum $L_3$ and corresponds to the eigenvalue $L_3 = N$. Hence they are similar to the GCS with conserved charge, examined recently by Skagerstam [10].

From formula (41) it immediately follows that the probability density reaches a maximum at a distance $r = (N/4\operatorname{Re} a)^{1/2}$, i.e. , it is proportional to the amplitude of the classical oscillator (30). It is easy to see also that the trajectory in phase space $D$ is again representable in forms (29), (36), i.e., as $SU(1,1)$ transformation, depending on the motion of classical oscillator (30).

Using the explicit expression (41) and usual quantum-mechanical rules one can compute the mean values of canonical operators, as follows [6]

$$\langle x^2 \rangle_{kz} = \frac{N+1}{4\operatorname{Re} a} = \frac{\rho^2}{\lambda_s} \frac{N+1}{2}, \quad \lambda_s = \frac{1 - s^*s}{|1-s|^2}, \qquad (42)$$
$$\langle p_x^2 \rangle_{kz} = \operatorname{Re} a \left(1 + (N+1)\frac{\operatorname{Im}^2 a}{\operatorname{Re}^2 a}\right) = \frac{\lambda_s}{2\rho^2}\left(1 + (N+1)\frac{4\operatorname{Im}^2 s}{(1 - s^*s)^2}\right).$$

The same expressions hold for $\langle y^2 \rangle_{kz}$ and $\langle p_y^2 \rangle_{kz}$. In the stationary case ($\rho = (2/\omega)^{1/2}$, $\gamma = \omega t/2$) these quantities vary in time according to usual classical equations of motion.

Let us choose in the moment $t = 0$ the initial conditions for Eq. (30) in such a manner that $s(0) = z$. Then the mean values (42) in the states $|k; z, 0\rangle$ ($k = (N+1)/2$) with real $z$, $\operatorname{Im} z = 0$, become

$$\langle x^2 \rangle_{kz} = \frac{\rho^2(1-z)}{1+z}\frac{N+1}{2},$$
$$\langle p_x^2 \rangle_{kz} = \frac{1+z}{2\rho^2(1-z)} \qquad (43)$$

Obviously the Heisenberg uncertainty product $\langle x^2 \rangle \langle p_x^2 \rangle = (N+1)/4$ preserves its value when the time in the disc $D$ is concentrated on the real axis. The states $|1/2; z\rangle$ with $\operatorname{Im} z = 0$ are the only MUS. More explicitly they can be written in the form

$$|1/2; z> = (1 - z^2)^{1/2} \sum_{n=0}^{\infty} z^n |n, n\rangle. \qquad (44)$$

Let us find now the mean value of the operator $\tilde{A} = A \otimes I$ in states (44) for $z = \exp(-\beta\omega/2)$, $\beta = 1/k_o T$, $k_o$ being the Boltzmann constant. The operator $A$ acts in Fock space spanned by the canonical basis $|n\rangle$. We have [7]

$$\langle 1/2; z|\tilde{A}|1/2; z\rangle = \left(1 - e^{-\beta\omega}\right) \sum_{n=0}^{\infty} e^{-\beta\omega n} \langle n|A|n\rangle = Z^{-1} \operatorname{tr}\left(A e^{-\beta H_0}\right),$$
$$Z = \operatorname{tr}\left(e^{-\beta H_0}\right), \quad H_0 = \omega a^\dagger a. \qquad (45)$$

According to Ref. 11 the states (44) may be considered as ground states of thermodynamical system, corresponding to different temperatures. Quite analogically one can construct the excited states $|N, \beta\rangle = (N!)^{-1/2} a^\dagger(\beta)|0, \beta\rangle$, $|0, \beta\rangle = |1/2; z\rangle$, where $a^\dagger(\beta) = U(\beta) a^\dagger U^{-1}(\beta)$, $|0, \beta\rangle = U(\beta)|0\rangle$. The states $|N, \beta\rangle$ form a real subsystem of the $SU(1,1)$ CS related to UIR $V_{(N+1)/2}^{(+)}$. The possibility of expressing the quantum-statistical averages (45) as usual quantum-mechanical mean values was, apparently, first observed by Y. Takahashi and H. Umezawa [11] without any connection to the theory of GCS. Here we established that the states of thermodynamical systems in thermostat are $SU(1,1)$ CS in the extended Hilbert space $\boldsymbol{H} \times \boldsymbol{H}$.

---

[6] Here $\langle x \rangle_{kz} = \langle y \rangle_{kz} = 0$. (Note added).

[7] More precisely $A|n, m; t\rangle = \sqrt{n}|n, m; t\rangle$. In (44) $m = n$ (i.e. $N = 0$). For stationary magnetic field $A = ae^{i\omega t}$ where $a$ does not depend on $t$ explicitly. (Note added).

## 5  U(N+1) Coherent States

Let us consider the $(N+1)$-level system, described by the Hamiltonian

$$H = \sum_{i,j=0}^{N} h_{ij}(t) a_i^\dagger a_j \tag{46}$$

where $h_{ij}^* = h_{ji}$ (Hermitian condition) and $[a_i, a^\dagger] = \delta_{ij}$. The Hamiltonian (46) belongs to the (ladder) representation of Lie algebra $u(N+1)$ spanned by the generators [7]

$$E_{ij}^+ = a_i^\dagger a_j \, (i > j), \quad E_{ij}^- = a_i^\dagger a_j \, (i > j), \quad E_{ij} = a_i^\dagger a_i \quad (i, j = 0, 1, 2, \ldots, N). \tag{47}$$

The UIR of $U(N+1)$ are determined by the highest weights $(m_0, m_1, \ldots, m_N)$, $m_0 \geq m_1 \geq \ldots \geq m_N$, where $m_0, m_1, \ldots, m_N$ are integers.

We shall construct GCS in the carrier space of the UIR $(m, 0, \ldots, 0)$ spanned by the vectors $|m_0, m1, \ldots, m_N\rangle = |m_0\rangle|m1\rangle \ldots |m_N\rangle$, where $m_0 + m1 + \ldots + m_N = m$ and $|m_i\rangle = (m_i!)^{-1/2} a_{ii}^{\dagger m}|0\rangle$. The fiducial vector is chosen to coincide with the weight vector $|m, 0, \ldots, 0\rangle$. Using a suitable parametrization of the group element we obtain the $U(N+1)$ CS in the form

$$\begin{aligned}|m; z\rangle &= C_z \exp\left(\sum_{i>0} z_i E_{i0}^+\right) |m, 0, \ldots, 0\rangle \\ &= (1 + z^* z)^{-m/2} \sum_{m_0 + \ldots + m_N = m} \frac{z_1^{m_1} \ldots z_N^{m_N} \sqrt{m!}}{\sqrt{m_0! \ldots m_N!}} |m_0, \ldots, m_N\rangle \ ,\end{aligned} \tag{48}$$

where $z^* z = \sum_{i=1}^N z_i^* z_i$. It is seen that in this case the role of the phase space is played by the $N$-dimensional complex space $C^N$. The states (48) obviously are generalization of the known $SU(2)$ and $SU(3)$ CS, the latter being constructed in Ref. 12.

Let us turn to the dynamics of $U(N+1)$ CS. Recall that according to general theory [1] the (stable) time evolution of GCS is described by the same formula (48) where one ought to replace $z_i$ by time-dependent functions, determined as solutions of Euler equations. By means of the scalar product

$$\langle m; y | m; z \rangle = (1 + y^* y)^{-m/2} (1 + z^* z)^{-m/2} (1 + y^* z)^m \tag{49}$$

one can obtain the Lagrangian (see (1))

$$\mathcal{L} = (im/2)(z^* \dot{z} - \dot{z}^* z)(1 + z^* z)^{-1} - \mathcal{H}, \tag{50}$$

where $\mathcal{H} = \langle m; z|H|m; z\rangle$. Hence the Euler equations have the form

$$\dot{z}_i + (\dot{z}_i z_j - z_i \dot{z}_j)z_j = (im)^{-1}(1 + z^*z)^2 \frac{\partial \mathcal{H}}{\partial z_j^*}. \tag{51}$$

For $N = 1$ Eqs. (51) are reduced to Eq. (7), related to $SU(2)$. Finally using the CS representation one can derive the following formula for the mean value of $a_i^\dagger a_j$ in $U(N+1)$ CS:

$$\langle a_i^\dagger a_j \rangle = (m/2)z_i^* z_j/(1 + z^*z). \tag{52}$$

Since the classical Hamiltonian $\mathcal{H}$ is a linear combination of such expressions we see that Eqs. (51) do not depend on the representation as in all the previous cases.

---

[8] previous e-print quant-ph/0407260.

Николов Б.А., Трифонов Д.А.            Е2-81-798
**Динамика обобщенных когерентных состояний.**
**II. Классические уравнения движения**

При использовании подхода Клаудера к стабильной эволюции обобщенных когерентных состояний рассмотрена эволюция этих состояний для групп $SU(2)$, $SU(1,1)$ и $SU(N)$. Показано, что одна и та же классическая функция $z(t)$ корректно описывает квантовую эволюцию для многих систем, как эквивалентных, так и неэквивалентных. Подробнее рассмотрены обобщенные когерентные состояния для сингулярного осциллятора с трением и для заряда в магнитном поле и показано, что эти состояния обладают стабильной эволюцией $SU(1,1)$-когерентных состояний. Квантовая динамика этих систем задается одним и тем же классическим решением $z(t)$. Рассмотрены физические свойства $SU(1,1)$-когерентных состояний и показано, что в случае дискретной серии $D_k^+$ они совпадают с состоянями, для которых квантовые средние совпадают со статистическими для осциллтора в термостате.